\documentclass[aps,pre,reprint,amsmath,amssymb,superscriptaddress,longbibliography,floatfix]{revtex4-2}

\usepackage[utf8]{inputenc}
\usepackage[english]{babel}

\usepackage{amsmath,amssymb,amsfonts,amsthm,mathrsfs,amsopn}
\usepackage{mathtools}
\usepackage[percent]{overpic}
\usepackage{graphicx}
\usepackage{xcolor}
\usepackage{bm}
\usepackage{braket}
\usepackage{pifont}
\usepackage{blkarray,bigstrut}
\usepackage{multirow}
\usepackage{array}
\usepackage{booktabs}
\usepackage{dcolumn}
\usepackage{float}
\usepackage{svg}
\usepackage{comment}
\usepackage{verbatim}

\usepackage{hyperref}

\makeatletter
\renewcommand{\[}{\begin{equation}}
\renewcommand{\]}{\end{equation}}
\makeatother

\def\be{\begin{equation}}
\def\ee{\end{equation}}
\def\bc{\begin{center}}
\def\ec{\end{center}}
\def\bea{\begin{eqnarray}}
\def\eea{\end{eqnarray}}

\theoremstyle{plain}

\begin{document}

\title{Delay-induced multistability in one-dimensional swarmalators with common intrinsic frequency}

\author{Rommel Tchinda Djeudjo}
\email{rommel.tchindadjeudjo@unamur.be}
\affiliation{Department of Mathematics \& naXys, Namur Institute for Complex Systems, University of Namur, Rue Grafé 2, B5000 Namur, Belgium}

\author{Timoteo Carletti}
\email{timoteo.carletti@unamur.be}
\affiliation{Department of Mathematics \& naXys, Namur Institute for Complex Systems, University of Namur, Rue Grafé 2, B5000 Namur, Belgium}

\author{K. P. O'Keeffe}
 \email{kevin.p.okeeffe@gmail.com}
 \affiliation{Starling Research Institute, Seattle, WA 98112, USA}

\begin{abstract}
We study the delayed one-dimensional swarmalator model when all units share a common intrinsic frequency $\omega$. Unlike the zero-frequency case, $\omega$ cannot be removed by a rotating-frame transformation because delayed interactions retain the phase accumulated over the lag. The result is a two-order-parameter analogue of the delayed Kuramoto model: the asynchronous state acquires incoherence lobes, the synchronized state becomes a rotating branch stable even when both the spatial and phase couplings are repulsive, and the phase wave develops narrow resonant stability bands. These delay-selected windows overlap to produce broad branch coexistence, including a narrow window near $K=-J$ where all three canonical states---asynchronous, phase wave, and synchronized---are simultaneously stable. In regions where no canonical branch is stable, the order parameters oscillate persistently.
\end{abstract}

\maketitle

\section{Introduction}
Swarmalators are mobile oscillators whose spatial motion and phase synchronization are coupled \cite{o2022collective,yoon2022sync}. They model systems where assembly and synchrony feed back on one another, with applications ranging from active and biological matter \cite{quillen2021metachronal,yan2012linking,tsiairis2016self,riedl2023synchronization,riedel2005self,creppy2016symmetry} to robotic swarms and human collective motion \cite{barcis2019robots,barcis2020sandsbots,xu2026navigation,toiviainen2025modeling}. The one-dimensional ring model is especially tractable: the change of variables $\xi=x+\theta$, $\eta=x-\theta$ returns a solvable mean-field structure that supports exactly three collective states --- asynchronous, phase wave, and synchronized --- with known exact stability boundaries \cite{o2022collective,o2025stability,global_sync}. This tractability has made it a test bed for coupling disorder \cite{o2022swarmalators,hao2023attractive}, phase lag and frustration \cite{lizarraga2023synchronization,senthamizhan2025frustration,sharma2025phaseLag,sharma2026forcedPhaseLag,sharma2026resonantPhaseLag}, directed interactions \cite{yu2025directed}, pinning and forcing \cite{sar2023pinning,sar2024solvable,sar2023swarmalators,anwar2024forced,anwar2025forced}, noise \cite{hong2023swarmalators}, finite-range and topological coupling \cite{sar2025effects,gou2026topological}, higher-order interactions \cite{anwar2024collective,anwar2025twoDimHigherOrder}, inertia \cite{okeeffe2026inertia}, self-propulsion \cite{okeeffe2026selfpropulsion}, time delay \cite{okeeffe2026delay}, and related variants \cite{ghosh2025dynamics,ghosh2026emergent,Senthamizhan2026FrequencyWeighted,louodop2025topological,Acharya2025WinfreeSwarmalator,o2024solvable,schilcher2025multicircular,lee2025circular,kongni2025attractive,sar2026interplay}.

Time delay arises naturally when signals propagate with finite speed. Delay-coupled swarmalators have been studied in higher-dimensional settings \cite{blum2024swarmalators,Lambu2026DelayExplosive,kumpeerakij2025aging}, but exact stability calculations are difficult there. In the delayed one-dimensional model with zero natural frequencies, analytic progress is possible and the three canonical states persist with delay-modified boundaries \cite{okeeffe2026delay}; asymmetric delays in the spatial and phase channels have also been studied \cite{djeudjo2026asymmetric}. Here we ask what happens when all swarmalators share the same nonzero intrinsic frequency $\omega$. Without delay, $\omega$ is removable by changing coordinates, for the internal variable, to a rotating frame. With delay, it does not hold true in general: the state at time $t$ couples to the state at time $t-\tau$, so a uniform rotation leaves a phase offset $\omega\tau$ that acts as a delay-induced lag, if $\omega\tau$ is not a multiple of $2\pi$. This is precisely the mechanism behind the incoherence lobes in the delayed Kuramoto model \cite{yeung1999time}, but the swarmalator case is not a one-order-parameter copy of that story. The $\xi$--$\eta$ decomposition causes $\omega$ to enter the two coordinates with opposite signs, creating opposite delay phases in two coupled channels. The result is a richer diagram: the asynchronous state acquires incoherence lobes, the synchronized state becomes a rotating branch with delay-selected windows (including a regime where both couplings are repulsive yet delay restores synchrony), and the phase wave develops narrow resonant stability bands. These windows overlap, producing broad branch coexistence absent from both the nondelayed common-frequency model and the zero-frequency delayed model. The overlap is maximal in a narrow window near $K=-J$, where all three canonical branches are simultaneously linearly stable and each is realized as an attractor in simulation---a genuine \emph{tristability} of asynchronous, phase-wave, and synchronized states with no counterpart in either limiting model.

We derive exact closed-form stability boundaries for the asynchronous and synchronized branches, reduce the phase wave to an exact scalar characteristic equation evaluated over its numerically enumerated rotating branches, and characterize the leftover regions numerically. Throughout, asynchronous stability refers to damping of the first Fourier modes that generate the order parameters, $r$ and $s$ --- what we call \emph{active-mode} stability --- not convergence of every neutral higher harmonic. Table~\ref{tab:omega-comparison} summarizes how $\omega\neq0$ changes each branch relative to the $\omega=0$ delayed model.

\begin{table*}[tb]
\caption{How common intrinsic frequency changes the delayed one-dimensional swarmalator problem.}
\label{tab:omega-comparison}
\begin{ruledtabular}
\begin{tabular}{lcc}
feature & $\omega=0$ delayed model & common $\omega\neq0$ delayed model\\
\hline
asynchronous & negative-$K$ strip & incoherence lobes\\
synchronized & static locked branch & rotating branch with collective frequency $\Omega$; $\cos(\Omega\tau)$ windows\\
phase wave & locked/uniform branch & resonant locked/uniform branch\\
coexistence & delay-destabilized static branches & lobe-induced competition; tristable point\\
leftover regions & delay-destabilized; nonstatic & gaps between stable branches; unsteady where tested\\
\end{tabular}
\end{ruledtabular}
\end{table*}

\section{Model}
We consider $N$ swarmalators on a ring, with position $x_i\in S^1$ and phase $\theta_i\in S^1$. Let $x_{j,\tau}=x_j(t-\tau)$ and $\theta_{j,\tau}=\theta_j(t-\tau)$. The delayed original-coordinate model is
\begin{align}
\dot x_i
&=
\nu_i'
+\frac{J'}{N}\sum_j
\sin(x_{j,\tau}-x_i)\cos(\theta_{j,\tau}-\theta_i),
\label{eq:original-x}\\
\dot \theta_i
&=
\omega_i'
+\frac{K'}{N}\sum_j
\sin(\theta_{j,\tau}-\theta_i)\cos(x_{j,\tau}-x_i).
\label{eq:original-theta}
\end{align}
We set the natural spatial velocity to zero, $\nu_i'=0$, and take a common intrinsic phase frequency, $\omega_i'=\omega$. As already stated, let us assume  $\omega\tau\not\in 2\pi\mathbb{Z}$, otherwise Eqs.~\eqref{eq:original-x} and~\eqref{eq:original-theta} will be invariant once read in the rotating frame with velocity $\omega t$. Passing to the sum and difference coordinates
\begin{align}
\xi_i=x_i+\theta_i,\qquad \eta_i=x_i-\theta_i,
\end{align}
and using $\sin a\cos b=\tfrac12[\sin(a+b)+\sin(a-b)]$ gives the transformed system below, with
\begin{equation}
\label{eq:KJ}
K=\frac{J'+K'}{2},\qquad J=\frac{J'-K'}{2}.
\end{equation}
We define order parameters
\begin{align}
r e^{i\phi}=\left\langle e^{i\xi}\right\rangle,\qquad
s e^{i\psi}=\left\langle e^{i\eta}\right\rangle,
\end{align}
where $\langle\cdot\rangle=N^{-1}\sum_j$. The delayed mean fields are evaluated at $t-\tau$; for example $r_\tau=r(t-\tau)$ and $\phi_\tau=\phi(t-\tau)$.
The sums include delayed self-feedback. Its finite-$N$ contribution is $O(1/N)$ and therefore vanishes in the continuum stability calculations; retaining it keeps the finite-$N$ numerics in the same mean-field convention as the order parameters.

The equations used in the analysis are therefore
\begin{align}
\dot{\xi}_i
&=\omega
+K r_\tau\sin(\phi_\tau-\xi_i)
+J s_\tau\sin(\psi_\tau-\eta_i),
\label{eq:xi-model}\\
\dot{\eta}_i
&=-\omega
+J r_\tau\sin(\phi_\tau-\xi_i)
+K s_\tau\sin(\psi_\tau-\eta_i).
\label{eq:eta-model}
\end{align}
Here $K$ is the same-coordinate coupling and $J$ is the cross-coordinate coupling. Equivalently $K+J=J'$ and $K-J=K'$, so $K$ is the mean of the original spatial and phase couplings $J$ is their half-difference (see Eq.~\eqref{eq:KJ}); this lets us read the transformed-coordinate results back into the physical couplings (Sec.~\ref{sec:analysis}). For $J>0$, rescaling time by $J$ leaves only the ratios $K/J$, $\omega/J$, and the scaled delay $J\tau$. We therefore set $J=1$ in the figures, as in the baseline phase diagrams, and use $\omega=\pi/2$ as a representative value. However, the analytic formulas below keep $J$ and $\omega$ explicit.

\section{Numerics}
\label{sec:numerics}
\begin{figure*}[!t]
\centering
\includegraphics[width=0.92\textwidth]{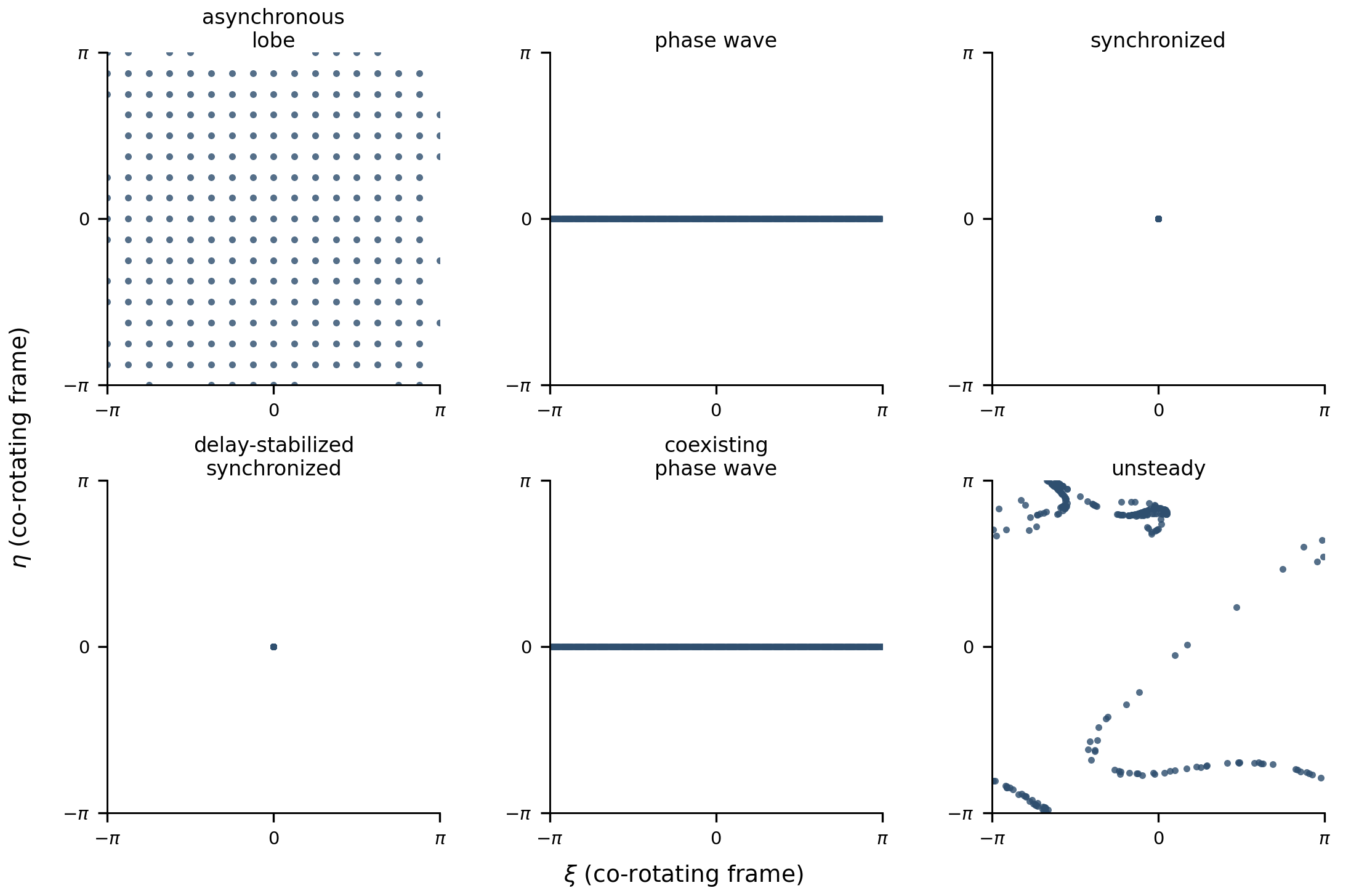}
\caption{Representative states for the identical-frequency delayed model in the $(\xi,\eta)$ plane. Each panel is shown in an appropriate co-moving frame: for a branch with frequencies $(\Omega_\xi,\Omega_\eta)$ we plot $(\xi_i-\Omega_\xi t,\eta_i-\Omega_\eta t)$ modulo $2\pi$. The first row shows the asynchronous lobe $(K,\tau)=(0.90,1.60)$, the phase wave $(0.50,0.10)$, and the synchronized state $(2.00,8.00)$. The second row shows a delay-stabilized synchronized state with negative same-coordinate coupling $(-1.60,0.50)$, a phase wave in a phase-wave/synchronized coexistence region $(1.05,7.00)$, and an unsteady trajectory at a point with no stable canonical branch $(0.50,0.50)$. Here $J=1$ and $\omega=\pi/2$. All states except the delay-stabilized synchronized panel are reproduced from independent random initial histories; the delay-stabilized synchronized state at $(K,\tau)=(-1.60,0.50)$ is multistable and requires a history prepared close to the synchronized branch (see Table~\ref{tab:identical-omega-robustness}).}
\label{fig:identical-omega-gallery}
\end{figure*}

We integrated Eqs.~\eqref{eq:xi-model}--\eqref{eq:eta-model} with a fourth-order Runge--Kutta method and a method-of-steps delay history. Unless otherwise stated, simulations used $N=256$, target step size $dt=0.03$, and total time $T=120$--$240$, depending on the delay. 
For the rotating synchronized and phase-wave branches, the history on $[-\tau,0]$ was initialized on the exact rotating solution; this avoids introducing an artificial delay mismatch at $t=0$. Random-history checks used independent uniform samples on the $(\xi,\eta)$ torus. Late-time statistics were computed over the final $35\%$ of each run. A trajectory was classified as unsteady if either order-parameter standard deviation exceeded $0.035$; otherwise, the mean values of $r$ and $s$ separated asynchronous, phase-wave, and synchronized states. Fig. \ref{fig:identical-omega-gallery} shows representative finite-$N$ states in the co-rotating frame.

We found the following states:

(i) \emph{Asynchronous state.} Both transformed coordinates are incoherent, so $r=s=0$. Unlike the zero-frequency case, the asynchronous state can be stable for positive $K$ inside finite-delay lobes. The particles drift in the laboratory transformed coordinates with velocities $\pm\omega$, but the distribution is stationary in the co-moving frame.

(ii) \emph{Phase wave.} One transformed coordinate is uniformly distributed while the other is locked. For the orientation shown in Fig.~\ref{fig:identical-omega-gallery}, $r=0$ and $s=1$ in a frame rotating with the branch frequencies $(\Omega_\xi,\Omega_\eta)$. Delay creates several phase-wave stability bands, including high-delay bands not present when $\omega=0$.

(iii) \emph{Synchronized state.} Both transformed coordinates are locked, so $r=s=1$ in a rotating frame. The ordinary attractive synchronized branch persists, but delay also stabilizes synchronized states for some negative $K$ values when the accumulated phase makes the effective linear coupling positive.

\begin{figure}[tb]
\centering
\includegraphics[width=\columnwidth]{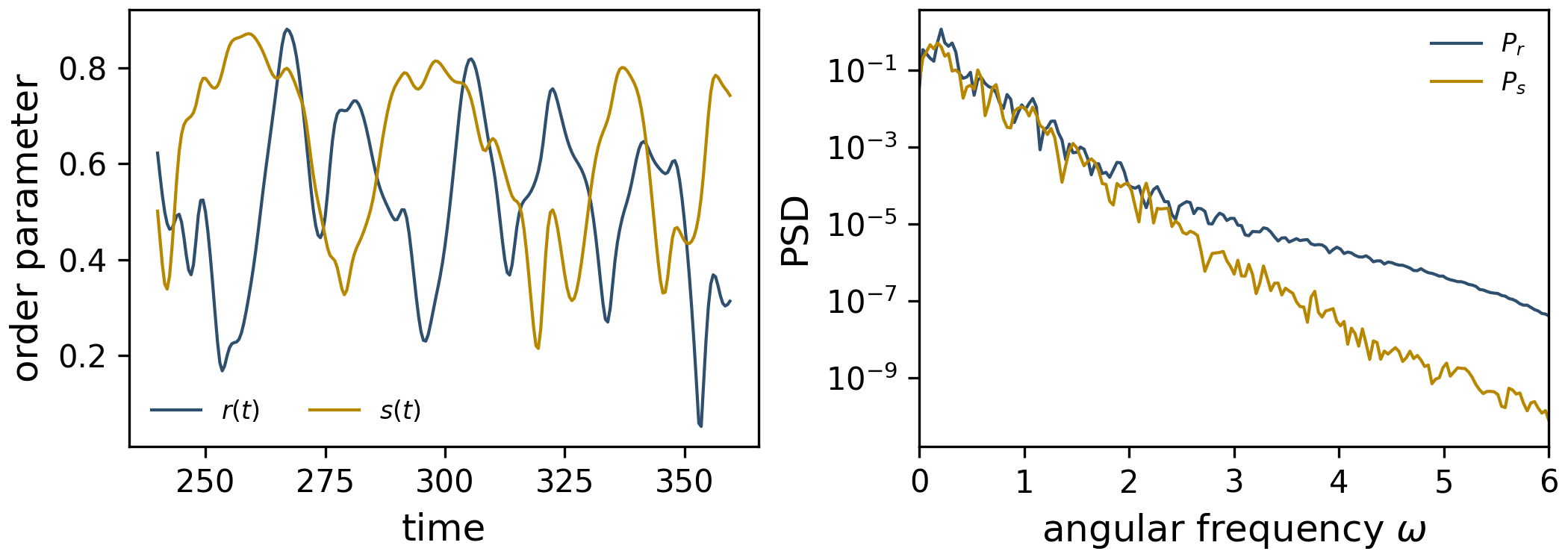}
\caption{Representative leftover point with no stable canonical branch, $(K,\tau)=(0.50,0.50)$ (same parameters as Fig.~\ref{fig:identical-omega-gallery}). Left: order-parameter time series $r(t)$, $s(t)$. Right: Welch power spectra of $r$ and $s$, showing sharp peaks at a dominant frequency and its harmonics (period $\approx30$) rather than broadband noise. The $0$-$1$ test of Gottwald and Melbourne \cite{gottwald2004} gives $\mathcal{K}_r\approx0.02$, $\mathcal{K}_s\approx0.03$, indicating regular (periodic/quasiperiodic) macroscopic dynamics, not chaos.}
\label{fig:identical-omega-unsteady-trace}
\end{figure}

(iv) \emph{Noncanonical unsteady dynamics.} At the tested points with no stable canonical branch, direct simulations show persistent order-parameter oscillations. We tested this with a basin survey: at three representative leftover points $(K,\tau)=(0.50,0.50)$, $(0.50,3.00)$, and $(-0.30,2.00)$, all $100$ random initial histories at each point settled to persistent oscillations---none to a static cluster or other attractor---and the classification was unchanged when the horizon was extended to $T=600$. Other states (static clusters, quasiperiodic branches, long transients) cannot be excluded in general; but where tested the unsteady label is robust. These oscillations are regular rather than chaotic: at the representative point the power spectra of $r$ and $s$ are sharply peaked at a dominant frequency and its harmonics, and the $0$-$1$ test for chaos \cite{gottwald2004} returns $\mathcal{K}_r,\mathcal{K}_s\lesssim0.03$ (Fig.~\ref{fig:identical-omega-unsteady-trace}), consistent with periodic or quasiperiodic macroscopic dynamics. The representative point in Fig.~\ref{fig:identical-omega-gallery} has persistently oscillating order parameters and no linearly stable canonical branch.

\begin{figure*}[!t]
\centering
\includegraphics[width=0.93\textwidth]{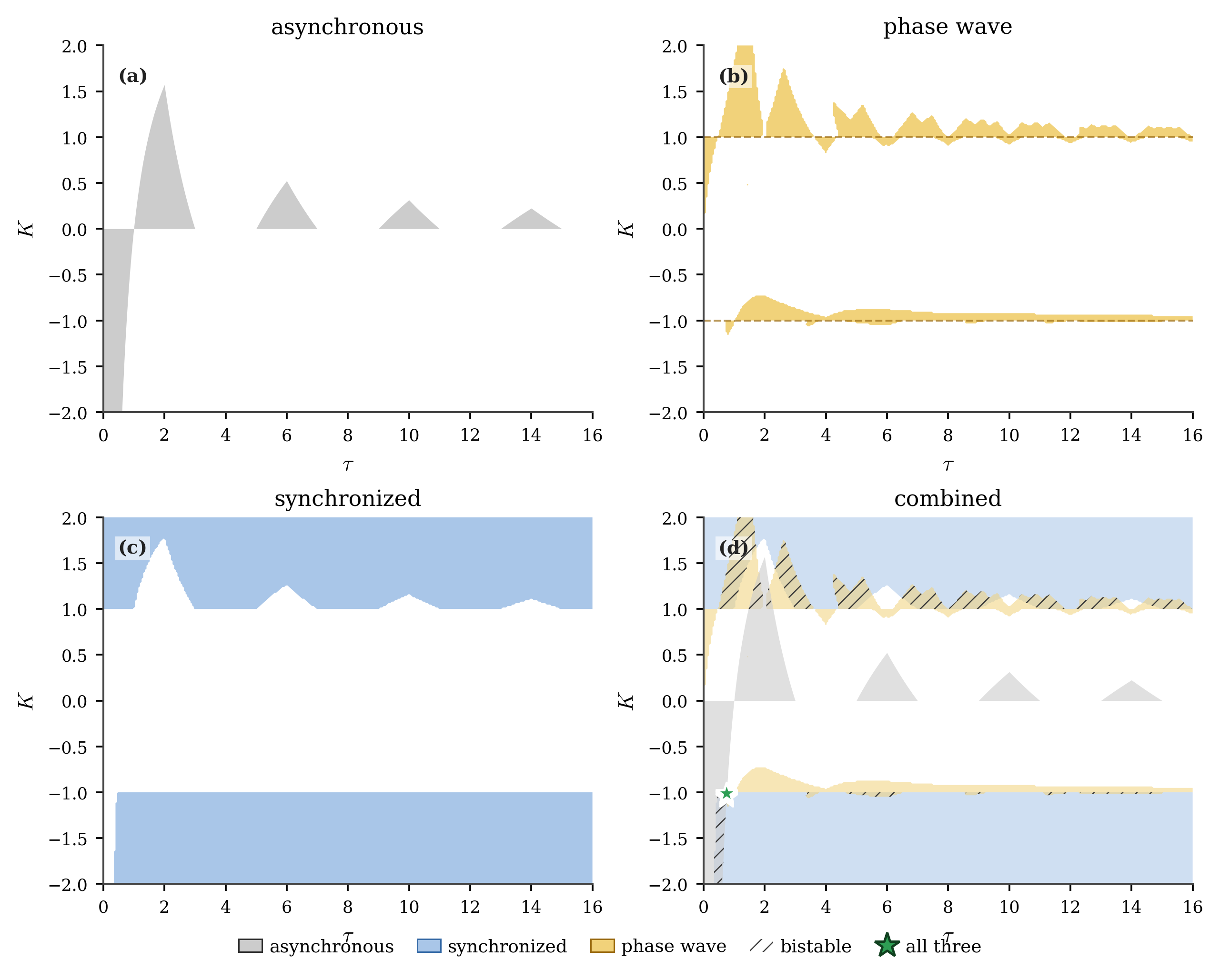}
\caption{Stability regions for $J=1$ and $\omega=\pi/2$. (a)~The asynchronous region (active-mode stability: damping of the first-harmonic order-parameter sector, neutral higher harmonics excluded) is the exact set of incoherence lobes of the delayed Kuramoto model [Eqs.~\eqref{eq:async-negative},~\eqref{eq:async-lobes}]. (b)~The phase-wave region (gold): a small region near $0<K<J$ at low delay plus thin bands that hug the exact real boundary $K=\pm J$ (dashed, from $\lambda=0$ in Eq.~\eqref{eq:pw-char-omega}); brown curves are representative analytic Hopf branches [Eqs.~\eqref{eq:pw-hopf-real-omega}--\eqref{eq:pw-quartic-omega}]. The region is cross-validated by argument-principle root counting (App.~\ref{app:pw-validation}). (c)~The synchronized stable region (blue): rotating branches satisfying Eq.~\eqref{eq:sync-stability}, which requires $\cos(\Omega\tau)>0$ for $K>J$ and $\cos(\Omega\tau)<0$ for $K<-J$; the marginal curves are given by Eq.~\eqref{eq:sync-boundaries}. (d)~Combined diagram (legend): the asynchronous (gray), synchronized (blue), and phase-wave (gold) stable regions overlaid; the hatched region is bistable (exactly two branches stable) and the green star marks the small set near $(K,\tau)\approx(-1.04,0.74)$ where all three are simultaneously stable. White regions have no linearly stable canonical branch; simulations there show persistent order-parameter oscillations (Sec.~\ref{sec:numerics}).}
\label{fig:identical-omega-stability}
\end{figure*}

\begin{figure*}[!t]
\centering
\includegraphics[width=0.87\textwidth]{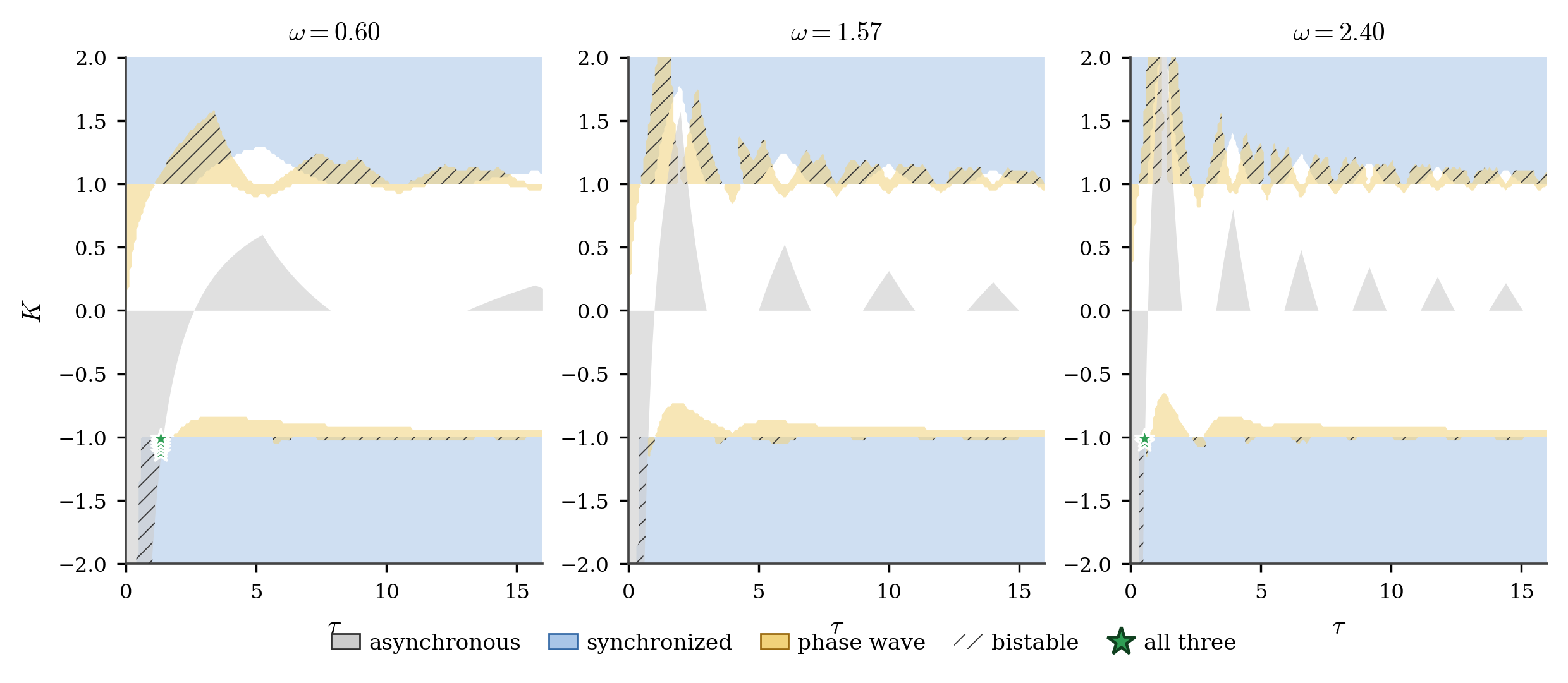}
\caption{Dependence of the combined stability diagram on the common frequency. The same color convention as Fig.~\ref{fig:identical-omega-stability} is used. Increasing $\omega$ shifts the positive-$K$ asynchronous lobes to smaller $\tau$ and narrows them in $\tau$, while raising their $K$ ceiling ($K<\omega/(2m-1)$) and fitting more lobes into the window; the delay-selected coexistence between canonical branches persists. These panels use the same stability tests as Fig.~\ref{fig:identical-omega-stability}.}
\label{fig:identical-omega-slices}
\end{figure*}

\begin{table*}[tb]
\caption{Representative finite-$N$ realizations for $J=1$ and $\omega=\pi/2$. The order-parameter means and standard deviations are computed over the final $35\%$ of each run.}
\label{tab:identical-omega-validation}
\begin{ruledtabular}
\begin{tabular}{lcccccc}
case & $K$ & $\tau$ & predicted stable & observed & $(\bar r,\bar s)$ & $(\sigma_r,\sigma_s)$\\
\hline
asynchronous lobe & $0.90$ & $1.60$ & asynchronous & asynchronous & $(0.000,0.000)$ & $(0.000,0.000)$\\
phase wave & $0.50$ & $0.10$ & phase wave & phase wave & $(0.000,1.000)$ & $(0.000,0.000)$\\
synchronized & $2.00$ & $8.00$ & synchronized & synchronized & $(1.000,1.000)$ & $(0.000,0.000)$\\
delay-stabilized synchronized & $-1.60$ & $0.50$ & asynchronous, synchronized & synchronized & $(1.000,1.000)$ & $(0.000,0.000)$\\
coexisting phase wave & $1.05$ & $7.00$ & phase wave, synchronized & phase wave & $(0.000,1.000)$ & $(0.000,0.000)$\\
noncanonical unsteady & $0.50$ & $0.50$ & none & unsteady & $(0.546,0.559)$ & $(0.167,0.200)$\\
\end{tabular}
\end{ruledtabular}
\end{table*}

\section{Analysis}
\label{sec:analysis}

\subsection{Asynchronous state}
The asynchronous state is the uniform density on the $(\xi,\eta)$ torus. In the co-moving frame, stability in the active order-parameter sector is controlled by the two first Fourier modes, which satisfy the same scalar characteristic equation up to complex conjugation:
\begin{align}
\lambda+i\omega=\frac{K}{2}e^{-\lambda\tau}.
\label{eq:async-char-omega}
\end{align}
The cross coupling $J$ drops out of this order-parameter sector because, about the uniform torus, it couples to mixed Fourier components orthogonal to the first harmonics that generate $r$ and $s$. Thus the active incoherence modes are controlled only by the same-coordinate coupling $K$.
All other Fourier modes are neutral in the order-parameter linearization, as in the zero-frequency model. Thus stability of the asynchronous state is understood here as \emph{active-mode} stability: linear damping of the nonneutral order-parameter modes, the neutral higher harmonics being excluded (as in the delayed Kuramoto analysis \cite{yeung1999time}). That this criterion captures the true asymptotic behavior is corroborated by the finite-$N$ basin survey (Table~\ref{tab:identical-omega-robustness}): random full-state histories at the lobe point converge to incoherence for every $N$ tested.

For $K<0$, the active mode is stable at $\tau=0$ and loses stability at the first Hopf crossing. Setting $\lambda=i\beta$ gives
\begin{align}
\tau<\frac{\pi}{2\omega-K}.
\label{eq:async-negative}
\end{align}
For $K>0$, the asynchronous state is unstable at zero delay but can be stable in finite-delay lobes. The lobe boundaries are obtained from $\beta\tau=-\pi/2,-3\pi/2,\ldots$. For $m=1,2,\ldots$,
\begin{align}
K<\frac{\omega}{2m-1},
\qquad
\frac{(4m-3)\pi}{2\omega-K}
<
\tau
<
\frac{(4m-1)\pi}{2\omega+K}.
\label{eq:async-lobes}
\end{align}
Thus the asynchronous stability region is the direct swarmalator analogue of the incoherence lobes of the delayed Kuramoto model \cite{yeung1999time}. In the limit $\omega\to0$, the positive-$K$ lobes collapse and Eq.~\eqref{eq:async-negative} reduces to the zero-frequency strip $-\pi/\tau<K<0$.

\subsection{Synchronized state}
The rotating synchronized state is
\begin{align}
\xi_i(t)=\Omega t+\xi_0,\qquad
\eta_i(t)=-\Omega t+\eta_0 .
\end{align}
Substituting into Eqs.~\eqref{eq:xi-model}--\eqref{eq:eta-model} gives the self-consistency equation
\begin{align}
\Omega=\omega+(J-K)\sin(\Omega\tau).
\label{eq:sync-omega-self}
\end{align}
For fixed $(K,\tau)$ this equation may have multiple rotating solutions. The synchronized region in Fig.~\ref{fig:identical-omega-stability} is the union over all real roots $q=\Omega\tau$ that satisfy the stability condition below.
Let
\begin{equation}
\label{eq:cdef}
q=\Omega\tau,\qquad c=\cos q.
\end{equation}
Write perturbations in the co-rotating frame as
$u_i=\delta\xi_i$ and $v_i=\delta\eta_i$. To linear order,
\begin{align}
\dot u_i
&=
c\left[
K(\bar u_\tau-u_i)+J(\bar v_\tau-v_i)
\right],
\label{eq:sync-linear-u}\\
\dot v_i
&=
c\left[
J(\bar u_\tau-u_i)+K(\bar v_\tau-v_i)
\right],
\label{eq:sync-linear-v}
\end{align}
where $\bar u=N^{-1}\sum_i u_i$, $\bar v=N^{-1}\sum_i v_i$, and the subscript $\tau$ denotes evaluation at $t-\tau$. Thus the symmetric and antisymmetric perturbations diagonalize the linearized dynamics with effective coefficients
\begin{align}
\mu_\pm=c(K\pm J).
\end{align}
The transverse modes have eigenvalues $-\mu_\pm$, while the collective delay modes satisfy
\begin{align}
\lambda+\mu_\pm(1-e^{-\lambda\tau})=0.
\label{eq:sync-delay-char}
\end{align}
Equation~\eqref{eq:sync-delay-char} always has the neutral root $\lambda=0$, corresponding to rigid shifts of the synchronized branch. If $\mu_\pm>0$, all other roots lie in the left half-plane: if a nonzero root had $\operatorname{Re}\lambda\geq0$, then
$\lambda=\mu_\pm(e^{-\lambda\tau}-1)$ would imply
$\operatorname{Re}\lambda\leq0$, with equality only at $\lambda=0$ (App.~\ref{app:sync-roots}).
Conversely, if either $\mu_+$ or $\mu_-$ is negative, the corresponding transverse eigenvalue $-\mu_\pm$ is positive. Hence the rotating synchronized branch is linearly stable, modulo the two neutral phase shifts, if and only if
\begin{align}
c(K+J)>0,\qquad c(K-J)>0.
\label{eq:sync-stability}
\end{align}
For $K>J$, this requires $c>0$ and recovers the ordinary attractive synchronized branch. For $K<-J$, delay can also stabilize synchronization when $c<0$. In the original couplings, where $K+J=J'$ and $K-J=K'$, the two conditions read $J'\cos(\Omega\tau)>0$ and $K'\cos(\Omega\tau)>0$: ordinary synchronization ($\cos\Omega\tau>0$) requires both the spatial and phase couplings to be attractive ($J',K'>0$), whereas the delay-stabilized $K<-J$ window is the striking regime in which \emph{both} original couplings are repulsive ($J',K'<0$) yet the accumulated delay phase ($\cos\Omega\tau<0$) makes the effective interaction attractive. This is a genuine physical effect of delay, not a transformed-coordinate artifact.

The marginal curves inside the synchronized regions occur when $c=0$, i.e.,
\begin{align}
q_n=\frac{\pi}{2}+n\pi.
\end{align}
Using Eq.~\eqref{eq:sync-omega-self}, these curves can be written explicitly as
\begin{align}
\tau_n(K)=
\frac{\frac{\pi}{2}+n\pi}
{\omega+(J-K)(-1)^n},
\label{eq:sync-boundaries}
\end{align}
Only branches with $\tau_n(K)>0$ are retained.
The straight boundaries $K=\pm J$ are the remaining synchronized-state stability boundaries, where one of the transverse coefficients $\mu_\pm$ vanishes.

\subsection{Phase wave}
Consider first the phase wave in which $\xi$ is uniformly distributed and $\eta$ is locked:
\begin{align}
\xi_i(t)=\Omega_\xi t+\frac{2\pi i}{N},\qquad
\eta_i(t)=\Omega_\eta t+\eta_0 .
\end{align}
Let
\begin{align}
q=\Omega_\eta\tau .
\end{align}
The locked coordinate satisfies
\begin{align}
q+\omega\tau+K\tau\sin q=0,
\label{eq:pw-q}
\end{align}
and the uniformly distributed coordinate rotates with
\begin{align}
\Omega_\xi=\omega-J\sin q.
\label{eq:pw-omega-xi}
\end{align}
Multiple phase-wave branches may coexist because Eq.~\eqref{eq:pw-q} can have multiple roots. The opposite orientation, in which $\xi$ is locked and $\eta$ is uniformly distributed, is obtained by the symmetry $\xi\leftrightarrow\eta$, $\omega\to-\omega$. Its locked-coordinate phase $q=\Omega_\xi\tau$ satisfies
\begin{align}
q-\omega\tau+K\tau\sin q=0,
\end{align}
and the uniformly distributed coordinate rotates with $\Omega_\eta=-\omega-J\sin q$. The stability test below is applied to both orientations when drawing Fig.~\ref{fig:identical-omega-stability}.

Perturbations that only displace the locked coordinate require
\begin{align}
Kc>0\, ,
\label{eq:pw-zeroth}
\end{align}
where $c$ is defined in~\eqref{eq:cdef}. The resonant first-Fourier sector gives the characteristic equation
\begin{align}
2\lambda^2+2Kc\lambda+
\left[c(J^2-K^2)-K\lambda\right]
e^{-(\lambda+i\Omega_\xi)\tau}=0.
\label{eq:pw-char-omega}
\end{align}
At zero delay, $q=0$ and Eq.~\eqref{eq:pw-char-omega} reduces to
\begin{align}
2\lambda^2+K\lambda+J^2-K^2=0,
\end{align}
so the phase wave is stable for
\begin{align}
0<K<J.
\end{align}

The Hopf boundary follows by setting $\lambda=i\nu$, $\nu>0$, and
\begin{align}
\Theta=(\nu+\Omega_\xi)\tau.
\end{align}
Separating Eq.~\eqref{eq:pw-char-omega} into real and imaginary parts gives
\begin{align}
-2\nu^2+c(J^2-K^2)\cos\Theta-K\nu\sin\Theta&=0,
\label{eq:pw-hopf-real-omega}\\
2Kc\nu-c(J^2-K^2)\sin\Theta-K\nu\cos\Theta&=0.
\label{eq:pw-hopf-imag-omega}
\end{align}
Eliminating $\Theta$ gives a quadratic equation in $\nu^2$:
\begin{align}
4\nu^4+K^2(4c^2-1)\nu^2-c^2(J^2-K^2)^2=0.
\label{eq:pw-quartic-omega}
\end{align}
The positive root gives the Hopf frequency explicitly:
\begin{align}
\nu^2 = \frac{K^2(1-4c^2)+\sqrt{K^4(4c^2-1)^2+16c^2(J^2-K^2)^2}}{8}.
\label{eq:pw-nu}
\end{align}
Equations~\eqref{eq:pw-q} and \eqref{eq:pw-hopf-real-omega}--\eqref{eq:pw-quartic-omega} determine the phase-wave Hopf curves that underlie the colored phase-wave regions in Fig.~\ref{fig:identical-omega-stability}. Thus the phase-wave stability problem is reduced analytically to branch equations and a scalar characteristic equation, although the plotted regions are obtained by enumerating the real rotating branches numerically. In practice, for the colored diagram we solved the locked-coordinate branch equation for all real rotating branches in the interval
$q\in[-\omega\tau-|K|\tau,\,-\omega\tau+|K|\tau]$
(located by sign-change bracketing on a fine $q$-mesh augmented with the locked-coordinate extrema $q=\tfrac{\pi}{2}+n\pi$ and the branch fold points where $1+K\tau\cos q=0$, refined by bisection, with duplicate roots merged at tolerance $10^{-6}$, so that narrow root pairs born at saddle-nodes are bracketed),
and checked the rightmost roots of Eq.~\eqref{eq:pw-char-omega} from a mesh of complex initial guesses. The phase-wave search in Fig.~\ref{fig:identical-omega-stability} was carried out over the full plotted coupling range, $|K|\leq2J$. A phase-wave point was colored stable only when at least one branch of either orientation satisfied $Kc>0$ and the rightmost-root search in the resonant sector returned negative real part; the Hopf equations above provide the analytic boundary conditions. We validate this search with an independent argument-principle root count (App.~\ref{app:pw-validation}): the two methods agree on $99.9\%$ of sampled cells, and the thin high-delay bands are confirmed as genuine narrow windows hugging $K=\pm J$ rather than numerical artifacts.

\section{Discussion}
The main result of this work is that a common intrinsic frequency fundamentally changes the delayed 1D swarmalator problem. It enters the sum and difference coordinates with opposite signs and acts as a delay-induced phase lag, so it reorganizes all three canonical branches at once---the asynchronous state acquires incoherence lobes, the synchronized state becomes a rotating branch stable not only for $K>J$ but also for $K<-J$ when $\cos(\Omega\tau)<0$, and the phase wave inherits two delay phases and develops narrow stability bands at small delay and in high-delay bands.

These delay-selected windows overlap: the $(\tau,K)$ diagram shows broad coexistence of linearly stable branches, with asynchronous/synchronized and phase-wave/synchronized overlap and small asynchronous/phase-wave pockets near the first lobe. From different initial histories these coexisting branches give genuine multistability (e.g.\ the negative-$K$ point of Table~\ref{tab:identical-omega-robustness}, reached as either asynchronous or synchronized).

The most striking instance is a \emph{triple} coexistence. In a narrow window near $K=-J$---for $J=1$, $\omega=\pi/2$ it spans $K\in[-1.11,-1.00]$, $\tau\in[0.73,0.76]$, centered at $(K,\tau)\approx(-1.04,0.74)$---the asynchronous lobe, the negative-$K$ delay-stabilized synchronized branch, and a phase-wave band hugging $K=-J$ are \emph{all} simultaneously linearly stable (the green star in Fig.~\ref{fig:identical-omega-stability}(d)). This is not merely an overlap of linear-stability regions: at the representative point $(K,\tau)=(-1.045,0.74)$ each of the three states is a genuine attractor. Prepared histories initialized on the asynchronous, synchronized, and phase-wave branches (with $10^{-4}$ noise) each return to their own branch, giving late-time $(r,s)=(0,0)$, $(1,1)$, and $(0,1)$ respectively, with negligible fluctuations. The triple coexistence is not fine-tuned to a single frequency: it persists over a range of common frequencies (for $J=1$, up to $\omega\approx2$), the window remaining anchored near $K=-J$ and shrinking in $\tau$ as $\omega$ increases. A population of identical swarmalators at a single coupling and delay can therefore be locked into incoherence, full synchrony, or a phase wave purely by its history---a delay-enabled tristability that has no counterpart in either the nondelayed common-frequency model or the zero-frequency delayed model.

This is a genuinely two-order-parameter version of the delayed Kuramoto model \cite{yeung1999time}, and the coexistence, the tristable point, and the negative-$K$ synchronization are absent from both the nondelayed common-frequency model and the zero-frequency delayed model.

In representative leftover regions, simulations show persistent order-parameter oscillations: a basin survey found all random histories at the tested leftover points settling to such dynamics, none reaching a static cluster or other attractor (Sec.~\ref{sec:numerics}). Clusters, quasiperiodic branches, or long transients cannot be excluded elsewhere. Adding more realism---distributed natural frequencies, delayed spatial coupling, or two spatial dimensions \cite{o2024solvable}---may also be fruitful.

\begin{acknowledgments}
We thank the members of the Starling Research Institute for helpful discussions.
\end{acknowledgments}

\appendix

\section{Phase-wave resonant sector}
\label{app:phase-wave}
Here we give the derivation of Eq.~\eqref{eq:pw-char-omega}. Consider the phase-wave orientation with
\begin{align}
\xi_i(t)=\Omega_\xi t+a_i+u_i(t),\qquad
\eta_i(t)=\Omega_\eta t+\eta_0+v_i(t),
\end{align}
where $a_i=2\pi i/N$ and $|u_i|,|v_i|\ll1$. The locked-coordinate mean field gives, to linear order,
\begin{align}
s_\tau\sin(\psi_\tau-\eta_i)
=-\sin q+c(\bar v_\tau-v_i),
\end{align}
where $q=\Omega_\eta\tau$, $c=\cos q$, and $\bar v=N^{-1}\sum_i v_i$. Perturbations of the uniform coordinate enter the mean field only through the resonant first Fourier amplitude
\begin{align}
U(t)=\frac{1}{N}\sum_j u_j(t)e^{i a_j}.
\end{align}
The locked-coordinate perturbation has the corresponding resonant amplitude
\begin{align}
V(t)=\frac{1}{N}\sum_j v_j(t)e^{i a_j}.
\end{align}
Nonresonant Fourier modes do not contribute to $r e^{i\phi}$ at linear order; they are neutral relabeling modes of the uniformly distributed coordinate or damped locked-coordinate modes controlled by $Kc$. The complex conjugate resonant sector gives the conjugate characteristic equation and therefore the same stability boundary.

After subtracting the rotating branch equations and collecting the resonant first-Fourier terms, the amplitudes satisfy
\begin{align}
\dot U
&=
\frac{K}{2}e^{-i\Omega_\xi\tau}U(t-\tau)
-Jc\,V,
\label{eq:app-u}\\
\dot V
&=
\frac{J}{2}e^{-i\Omega_\xi\tau}U(t-\tau)
-Kc\,V.
\label{eq:app-v}
\end{align}
Eliminating $V$ from Eqs.~\eqref{eq:app-u}--\eqref{eq:app-v} gives the scalar delay equation
\begin{align}
2\ddot U+2Kc\dot U+
\left[c(J^2-K^2)-K\frac{d}{dt}\right]
e^{-i\Omega_\xi\tau}U(t-\tau)=0 .
\end{align}
Substituting $U(t)\propto e^{\lambda t}$ gives
\begin{align}
2\lambda^2+2Kc\lambda+
\left[c(J^2-K^2)-K\lambda\right]
e^{-(\lambda+i\Omega_\xi)\tau}=0,
\end{align}
which is Eq.~\eqref{eq:pw-char-omega}. The opposite phase-wave orientation follows by $\xi\leftrightarrow\eta$ and $\omega\to-\omega$, as described in the main text.

\section{Stability of the rotating synchronized branch}
\label{app:sync-roots}
The collective delay modes of the synchronized branch satisfy Eq.~\eqref{eq:sync-delay-char}, $\lambda+\mu_\pm(1-e^{-\lambda\tau})=0$ with $\mu_\pm=c(K\pm J)$. Besides the neutral root $\lambda=0$ (a rigid phase shift), we show that if $\mu_\pm>0$ every root has $\mathrm{Re}\,\lambda<0$. Suppose $\lambda\neq0$ with $\mathrm{Re}\,\lambda\geq0$. Then $\lambda=\mu_\pm(e^{-\lambda\tau}-1)$; since $\tau>0$, $|e^{-\lambda\tau}|=e^{-\tau\,\mathrm{Re}\,\lambda}\leq1$, so $e^{-\lambda\tau}-1$ lies in the closed unit disk centered at $-1$ and has $\mathrm{Re}(e^{-\lambda\tau}-1)\leq0$. With $\mu_\pm>0$ this forces $\mathrm{Re}\,\lambda\leq0$, hence $\mathrm{Re}\,\lambda=0$. Writing $\lambda=i\beta$, the real part of $i\beta=\mu_\pm(e^{-i\beta\tau}-1)$ gives $0=\mu_\pm(\cos\beta\tau-1)$, so $\cos\beta\tau=1$, $e^{-i\beta\tau}=1$, and $\beta=0$. Thus $\lambda=0$ is the only root with $\mathrm{Re}\,\lambda\geq0$. The transverse eigenvalues $-\mu_\pm$ are damped when $\mu_\pm>0$, so the branch is linearly stable (modulo the two neutral shifts) iff $c(K+J)>0$ and $c(K-J)>0$, Eq.~\eqref{eq:sync-stability}.

\section{Validation of the phase-wave region}
\label{app:pw-validation}
The phase-wave regions in Fig.~\ref{fig:identical-omega-stability} are drawn by enumerating the real rotating branches [roots of Eq.~\eqref{eq:pw-q}] and locating the rightmost root of Eq.~\eqref{eq:pw-char-omega} from a mesh of complex initial guesses. To confirm that this search does not miss roots, we independently recount the right-half-plane zeros of Eq.~\eqref{eq:pw-char-omega} by the argument principle. Writing $D(\lambda)$ for the left-hand side of Eq.~\eqref{eq:pw-char-omega}, the number of unstable roots of a branch is
\begin{align}
N_{\rm RHP}=\frac{1}{2\pi i}\oint_{\partial\mathcal{R}}\frac{D'(\lambda)}{D(\lambda)}\,d\lambda,
\end{align}
the winding number of $D$ around a rectangle $\mathcal{R}=[\epsilon,R]\times[-M,M]$ that covers the right half-plane, with $\epsilon=10^{-6}$, $R=25$, $M=45$. These bounds enclose all unstable roots: for $\mathrm{Re}\,\lambda\geq0$ one has $|e^{-(\lambda+i\Omega_\xi)\tau}|=e^{-\tau\,\mathrm{Re}\,\lambda}\leq1$, so with $J=1$, $|c|\leq1$, and $|K|\leq2J$ over the plotted range,
\begin{eqnarray}
|D(\lambda)|&\geq& 2|\lambda|^2-|2Kc\,\lambda|-\big|c(J^2-K^2)-K\lambda\big|\\
&\geq &2|\lambda|^2-6|\lambda|-3\, ,
\end{eqnarray}
which is strictly positive for $|\lambda|>(3+\sqrt{15})/2\approx3.4$. Every right-half-plane zero therefore satisfies $|\lambda|\lesssim3.4$, far inside $\mathcal{R}$. A branch is stable iff $Kc>0$ and $N_{\rm RHP}=0$, and the phase wave is stable where at least one branch of either orientation is stable. This count independently validates the characteristic-root classification of each enumerated branch (completeness of the branch enumeration itself is handled by the fold-point subdivision of Eq.~\eqref{eq:pw-q} described in Sec.~\ref{sec:analysis}). On a coarse validation subgrid (every ninth $\tau$ and every seventh $K$ value, $1224$ cells) the argument-principle classification agrees with the seed-mesh classification used for Fig.~\ref{fig:identical-omega-stability} in $1223$ cells ($99.9\%$); the single discrepancy lies on a region boundary, where the rightmost root is within $\sim10^{-3}$ of the imaginary axis. The thin high-delay phase-wave bands are therefore genuine narrow stability windows hugging the real boundary $K=\pm J$, not numerical artifacts; their pixelated appearance reflects finite-grid sampling of a feature only a few cells wide.

\section{Numerical robustness, perturbation, and convergence checks}
\label{app:numerics}
This appendix collects the numerical validation supporting the stability boundaries and state classifications of the main text.

To avoid relying only on exact branch histories, we also perturbed the history on the entire interval $[-\tau,0]$ and measured direct perturbation norms. Here $u_i(t)$ and $v_i(t)$ denote deviations from the corresponding rotating branch in the co-moving frame. For the synchronized branch we remove the two neutral rigid shifts and use
\begin{align}
\Delta_{\rm sync}=
\left[
N^{-1}\sum_i\left\{(u_i-\bar u)^2+(v_i-\bar v)^2\right\}
\right]^{1/2}.
\end{align}
For the phase wave we use the active resonant mode of the uniformly distributed coordinate together with the locked-coordinate spread,
\begin{align}
\Delta_{\rm pw}
=
\left[
|U|^2+N^{-1}\sum_i(v_i-\bar v)^2
\right]^{1/2},\\
U=N^{-1}\sum_i u_i e^{ia_i},
\end{align}
where $a_i=2\pi i/N$ labels the uniformly distributed coordinate (App.~\ref{app:phase-wave}).
For the asynchronous state, the active perturbation is $\Delta_{\rm async}=(r^2+s^2)^{1/2}$. Table~\ref{tab:identical-omega-perturbations} reports the predicted non-neutral rightmost exponent from the corresponding characteristic equation, together with the measured initial and final perturbation norms. Stable representative branches decay, while nearby unstable-side branches grow. To confirm the integrator is converged near the stability boundaries, we also measured the growth/decay rate of the perturbation norm at three boundary-near branches for timesteps $dt$, $dt/2$, and $dt/4$ (Table~\ref{tab:identical-omega-convergence}). The measured rates are insensitive to $dt$ (confirming the integrator is converged) and agree with the predicted $\mathrm{Re}\,\lambda_{\max}$ in sign and---for the async and sync branches---in magnitude; the phase-wave magnitude is reduced by finite-amplitude saturation within its short pre-saturation window, not by timestep error. Table~\ref{tab:identical-omega-robustness} summarizes the robustness of the state labels to system size, timestep, and classification threshold.

\begin{table*}[tb]
\caption{Direct perturbation-norm checks around representative histories. The initial history is perturbed by independent Gaussian noise of amplitude $10^{-3}$ over $[-\tau,0]$. The exponent column gives the predicted non-neutral rightmost real part for the tested branch. The reported $\Delta_f$ is averaged over the final $20\%$ of the run.}
\label{tab:identical-omega-perturbations}
\begin{ruledtabular}
\begin{tabular}{lcccccc}
case & $K$ & $\tau$ & $\operatorname{Re}\lambda_{\max}$ & $\Delta_0$ & $\Delta_f$ & observed\\
\hline
asynchronous lobe & $0.90$ & $1.60$ & $-0.072$ & $4.32\times10^{-5}$ & $7.02\times10^{-9}$ & decays\\
asynchronous outside lobe & $0.90$ & $1.25$ & $0.054$ & $5.37\times10^{-5}$ & $2.35\times10^{-3}$ & grows\\
phase wave & $0.50$ & $0.10$ & $-0.048$ & $3.17\times10^{-4}$ & $4.97\times10^{-7}$ & decays\\
phase wave outside band & $0.50$ & $0.50$ & $0.177$ & $6.16\times10^{-4}$ & $1.30$ & grows\\
synchronized & $2.00$ & $8.00$ & $-0.724$ & $2.57\times10^{-4}$ & $<10^{-12}$ & decays\\
synchronized outside wedge & $0.80$ & $0.50$ & $0.130$ & $1.34\times10^{-3}$ & $1.16$ & grows\\
delay-stabilized synchronized & $-1.60$ & $0.50$ & $-0.238$ & $6.11\times10^{-4}$ & $4.72\times10^{-12}$ & decays\\
\end{tabular}
\end{ruledtabular}
\end{table*}

\begin{table}[tb]
\caption{Timestep convergence near stability boundaries. The measured perturbation-norm exponent (slope of $\log\|\delta\|$) at $dt=0.03$, $dt/2$, and $dt/4$ is compared with the predicted rightmost root $\mathrm{Re}\,\lambda_{\max}$. The measured rates are consistent in sign and within $15\%$ of the predicted magnitude; the async and sync magnitudes match closely, while the phase-wave magnitude is reduced by finite-amplitude saturation within its short pre-saturation fit window.}
\label{tab:identical-omega-convergence}
\begin{ruledtabular}
\begin{tabular}{lcccc}
branch & $\mathrm{Re}\,\lambda_{\max}$ & $dt$ & $dt/2$ & $dt/4$\\
\hline
async lobe (stable) & $-0.072$ & $-0.072$ & $-0.072$ & $-0.072$\\
phase wave (unstable) & $0.177$ & $0.163$ & $0.154$ & $0.160$\\
sync (unstable) & $0.130$ & $0.131$ & $0.130$ & $0.130$\\
\end{tabular}
\end{ruledtabular}
\end{table}

\begin{table*}[tb]
\caption{Robustness checks for the representative points in Fig.~\ref{fig:identical-omega-gallery}. Ordinary representative points are reached from random histories; the negative-$K$ synchronized point is multistable; halving $dt$ and varying the unsteady threshold by a factor of two preserve the labels.}
\label{tab:identical-omega-robustness}
\begin{ruledtabular}
\begin{tabular}{lccc}
case & check & result & late-time diagnostics\\
\hline
asynchronous lobe & random histories, $N=128,256,512$ & asynchronous for all $N$ & $\bar r,\bar s<10^{-3}$\\
phase wave & random history, $N=256$ & phase wave & one of $\bar r,\bar s$ near $1$\\
synchronized & random history, $N=256$ & synchronized & $(\bar r,\bar s)=(1.000,1.000)$\\
delay-stabilized synchronized & random vs. prepared histories & asynchronous vs. synchronized & confirms coexistence\\
noncanonical unsteady & random histories, $N=128,256,512$ & unsteady for all $N$ & $\sigma_r,\sigma_s=O(10^{-1})$\\
all rows & $dt\approx0.03$ vs. $dt\approx0.015$ & unchanged labels & same qualitative statistics\\
\end{tabular}
\end{ruledtabular}
\end{table*}

\section*{Data availability}
The simulation and analysis scripts that reproduce all figures and tables are openly available in a public repository, archived at Zenodo upon publication [DOI to be inserted].

\end{document}